\documentclass[twocolumn,english,aps,pra,showpacs]{revtex4}
\usepackage{graphicx}
\usepackage{amsfonts}
\usepackage{amsmath}
\usepackage{bm}
\usepackage{babel}

\newcommand{\picdir}{}

\begin{document}

\title{Hydrogen atom in crossed electric and magnetic fields: Phase space topology and torus quantization via periodic orbits }

\author{Stephan Gekle}
\author{J\"{o}rg Main}
\affiliation{Institut f\"{u}r Theoretische Physik 1, Universit\"{a}t Stuttgart,
70550 Stuttgart, Germany}

\author{Thomas Bartsch}
\affiliation{Department of Mathematical Sciences, Loughborough University, Loughborough LE11 3TU, UK}

\author{T. Uzer}
\affiliation{Center for Nonlinear Science, Georgia Institute of Technology, Atlanta,
Georgia 30332-0430, USA}

\date{\today}

\begin{abstract}
A hierarchical ordering is demonstrated for the periodic orbits in
a strongly coupled multidimensional Hamiltonian system, namely the
hydrogen atom in crossed electric and magnetic fields. It mirrors
the hierarchy of broken resonant tori and thereby allows one to characterize
the periodic orbits by a set of winding numbers. With this knowledge,
we construct the action variables as functions of the frequency ratios
and carry out a semiclassical torus quantization. The semiclassical energy levels thus obtained agree well with exact quantum calculations.
\end{abstract}


\maketitle

\section{Introduction}

The hydrogen atom in crossed electric and magnetic fields is among
the paradigmatic examples of strongly coupled multidimensional systems.
During the last two decades, a large number of experimental and theoretical
investigations (eg.,~\cite{Firstexperiment,QuasiLandau,2DsignificanceWalther,FloethmannKepler,FloethmannMoon,Turgayionization,
Experimentsymmetrybreak,Janionization,Wangplanar,Arnoldweb,Cushman,
Ericsonfluctuations,Ericsonfluctuations2})
have been devoted to the intricate physics of this seemingly elementary
nonintegrable system. In addition to its inherent interest, the hydrogen
atom in crossed fields can be used to study generic phenomena such
as Arnold diffusion \cite{Arnoldweb}, monodromy \cite{Cushman} or
Ericson fluctuations \cite{Ericsonfluctuations,Ericsonfluctuations2}.
Its deeper understanding also provides a solid basis to explore the
behavior of confined electrons in condensed matter physics, such as
in excitons \cite{SolidState1} and quantum dots \cite{Turgaysolidstate}.
Yet, in spite of the large amount of work spent upon it, the overall
phase space structure of the crossed-fields hydrogen atom still defies
a complete understanding. 

One of the most prominent features of a dynamical system are its periodic
orbits (POs). While individual POs can yield deep insights into the
local dynamics, in their entirety they provide an appropriate tool
to understand the global structure of a multidimensional phase space
\cite{Poincarebook}. Their knowledge is crucial in many applications
of classical dynamical systems such as astronomy \cite{Copenhagen,Jorba2,Poincarebook},
particle accelerators \cite{Resonance1} and fluid dynamics, e.g.~statistics
of turbulent flow \cite{Fluids1}. The semiclassical quantization
of the classical structures is an invaluable tool for the description
of molecular vibrations \cite{Joyeux97,Chemistry1,Vibrations1}, chemical
reactions \cite{Chemistry3} or the spectra of Rydberg atoms \cite{QuasiLandau,Wangplanar}. 

In this work we establish an organizing principle for POs in the crossed-fields
hydrogen atom. It allows the identification of winding numbers for
every PO that originates from the breakup of a torus. Conversely,
through this classification the POs elucidate the higher-dimensional
structures of the phase space and allow one to characterize their
topology. They enable us to reconstruct the hierarchy of broken tori
with sufficient precision to carry out an Einstein-Brillouin-Keller
(EBK) torus quantization \cite{Einstein,Brillouin1,Keller1,Percival77,Brack}.

The electron motion of a hydrogen atom exposed to an electric field
$F$ in the $x$ direction and a magnetic field $B$ in the $z$ direction,
is governed by the Hamiltonian, in atomic units \cite{Atomicunits},
\begin{equation}
H=\frac{1}{2}\mathbf{p}^{2}-\frac{1}{r}+\frac{B}{2}\left(p_{y}x-p_{x}y\right)+\frac{B^{2}}{8}\left(x^{2}+y^{2}\right)-Fx\;.
\label{eqn:Hamiltonian}
\end{equation}
Here $\mathbf{r}=(x,y,z)$ are the usual Cartesian coordinates, $\mathbf{p}=(p_x,p_y,p_z)$ the conjugate momenta, and $r=\sqrt{x^{2}+y^{2}+z^{2}}$. By virtue of the scaling properties
\cite{Friedrichscaling} of the Hamiltonian (\ref{eqn:Hamiltonian}),
if all classical quantities are multiplied by suitable powers of the
magnetic field strength, the dynamics can be shown not to depend on
the energy $E$ and the field strengths $B$ and $F$ separately,
but only on the scaled energy $\tilde{E}=B^{-2/3}E$ and the scaled
electric field strength $\tilde{F}=B^{-4/3}F$. In particular, coordinates
scale according to $\tilde{\mathbf{r}}=B^{2/3}\mathbf{r}$ and the
classical actions obey $\tilde S=B^{1/3}S$. In this paper, we present
results for a scaled electric field strength $\tilde{F}=0.5$ and
two scaled energies $\tilde{E}=-1.5$ and $\tilde{E}=-1.4$ slightly
below and slightly above the classical ionization threshold $\tilde{E}_{\text{I}}=-2\sqrt{\tilde{F}}$,
respectively. 

The organization of this paper is as follows. In Sec.~\ref{sec:review}
we give a brief general review of the phase space structures in Hamiltonian
systems that are relevant to our investigations. Sec.~\ref{sec:orga}
explains the dynamical principles underlying the organization of periodic
orbits and summarizes the organization of periodic orbits of the hydrogen
atom in crossed electric and magnetic fields. In Sec.~\ref{sec:WN}
we describe this organization in detail and show how it can be exploited
to assign winding numbers to the periodic orbits. Sec.~\ref{sec:tori}
presents the calculation of action variables from the information
thus obtained. A semiclassical Einstein-Brillouin-Keller quantization
is carried out in Sec.~\ref{sec:quant}, and conclusions are given
in Sec.~\ref{sec:conclusion}.

\section{Phase space structures in Hamiltonian systems}

\label{sec:review}

Among all Hamiltonian systems, those that exhibit the most regular
dynamics are the integrable systems, where almost every bounded trajectory
is confined to an invariant torus. At the other extreme, in ergodic
systems almost every trajectory comes arbitrarily close to every energetically
allowed point in phase space. In this work we treat the crossed-fields
hydrogen atom in a parameter range where the phase space structure
is dominated by the remnants of invariant tori. Nevertheless, the
external fields are chosen too strong for perturbation theory to be
a reliable tool, and we will consider the full nonintegrable Hamiltonian
(\ref{eqn:Hamiltonian}). 

By definition, a Hamiltonian system with $f$ degrees of freedom is
integrable if it possesses $f$ constants of motion in involution.
The regular level sets of these constants are $f$-dimensional tori
if they are compact \cite{Arnold,LiLi}. For singular values of the
constants, tori of lower dimensions can arise. Tori of different dimensions
form a hierarchy in which the tori of dimension $N$ are organized
around the tori of dimension $N-1$. Such hierarchies have been observed
in a variety of different systems (see, e.g., \cite{Joyeux97,PerturbationJan,Cushman}).
In the following sections, we will demonstrate how this hierarchy
of tori in an integrable limit can give rise to a hierarchy of isolated
POs in a nonintegrable system and how these POs can in turn be used
to gain information on the original hierarchy of invariant tori. 

In an integrable system, action-angle variables ($\mathbf{I}$, $\bm{\theta}$)
with the following properties can be introduced. The angles $\bm{\theta}$
determine the position on an individual torus. They increase linearly
with time \begin{equation}
\bm{\theta}=\bm{\omega}t+\bm{\theta}_{0},\label{eqn:frequencies}\end{equation}
 with a constant frequency vector $\bm{\omega}$ and initial conditions
$\bm{\theta}_{0}$. The conjugate action variables $\mathbf{I}$ are
constants of motion and characterize the invariant tori. They are
given by \begin{equation}
I_{i}=\frac{1}{2\pi}\oint_{\gamma_{i}}\mathbf{p}d\mathbf{q},\label{eqn:action}\end{equation}
 where $\gamma_{i}$ is the loop on the torus obtained as the angle
$\theta_{i}$ varies from 0 to $2\pi$ with all other angles held
fixed. 

The action variables~(\ref{eqn:action}) form the basis for a semiclassical
torus quantization \cite{Einstein,Brillouin1,Keller1,Percival77,Brack}. Finding
them explicitly, however, is a highly nontrivial task. Due to its
significance in applications such as the calculation of molecular
vibration spectra, the construction of the invariant tori has attracted
considerable attention \cite{Tannenbaum,Miller,Fourier1}. 

In an integrable system, the Hamiltonian can be written as a function
of the action variables only, and the frequencies $\bm{\omega}$ in
(\ref{eqn:frequencies}) are given by \begin{equation}
\omega_{i}=\frac{\partial H}{\partial I_{i}}.\end{equation}
 If the ratios of any two of the $\omega_{i}$ on a torus are rational,
that torus carries periodic orbits and is called a resonant torus.
In this case, a set $\mathbf{w}$ of integer winding numbers can be
found so that \begin{equation}
\omega_{1}:\omega_{2}:\ldots:\omega_{n}=w_{1}:w_{2}:\ldots:w_{n}.\label{eqn:ratios}\end{equation}
 Each winding number $w_{i}$ specifies the number of rotations along
the direction of the fundamental loop $\gamma_{i}$ that the POs execute
before repeating themselves. 

Angle coordinates $\bm{\theta}$ on a given torus can be defined in
various ways. Apart from a choice of origin, which is inconsequential,
any two angle coordinate systems are related by a linear transformation
\begin{equation}
\bm{\theta'}=M\cdot\bm{\theta}\end{equation}
 where $M$ is an $n\times n$ integer matrix with unit determinant
\cite{Borneng}. The action coordinates must be transformed according
to \begin{equation}
\mathbf{I'}=(M^{\text{T}})^{-1}\cdot\mathbf{I},\label{eqn:actionstrafo}\end{equation}
 whereas the winding numbers on a rational torus transform as
\begin{equation}
 \mathbf{w'}=M\cdot\mathbf{w}.
 \end{equation}
This freedom to choose different action-angle coordinate systems, together with the intimate connection between coordinate systems and winding numbers, will play a crucial role in our investigation of the phase space topology.

In a non-integrable system, a resonant torus breaks up into isolated
POs \cite{Arnold,PBtheorem,Meiss}. We use these POs as representatives
of the torus they stem from and call them $N$-torus POs, where $N$
is the dimension of the original torus. According to Kolmogorov-Arnold-Moser (KAM) theory \cite{Arnold,LiLi},
most nonresonant tori remain intact in a near-integrable system and break up only gradually. They
are interspersed with the isolated POs in the same way as resonant
and nonresonant tori are interspersed in an integrable system. Therefore,
POs can be used to investigate the structure of the surviving tori. 

According to the Poincar\'{e}-Birkhoff theorem \cite{PBtheorem,Arnold,LiLi},
a 2-torus in a two-degree-of-freedom system breaks up into an even
number of isolated POs, usually two. One of the POs is stable ({}``elliptic''),
the other is unstable ({}``hyperbolic''). Because both POs stem
from the same torus, their periods and actions differ only slightly. 

For systems with three degrees of freedom, Kook and Meiss \cite{Meiss}
were able to derive from symmetry considerations that a 3-torus breaks
up into four isolated POs that usually represent the four possible
stability combinations: one PO elliptic in both degrees of freedom
(ee), one PO hyperbolic in both degrees of freedom (hh) and two POs
elliptic in one degree of freedom and hyperbolic in the second (eh
and he).

\section{Organization of periodic orbits}
\label{sec:orga}

\begin{figure}
\includegraphics[width=.46\textwidth]{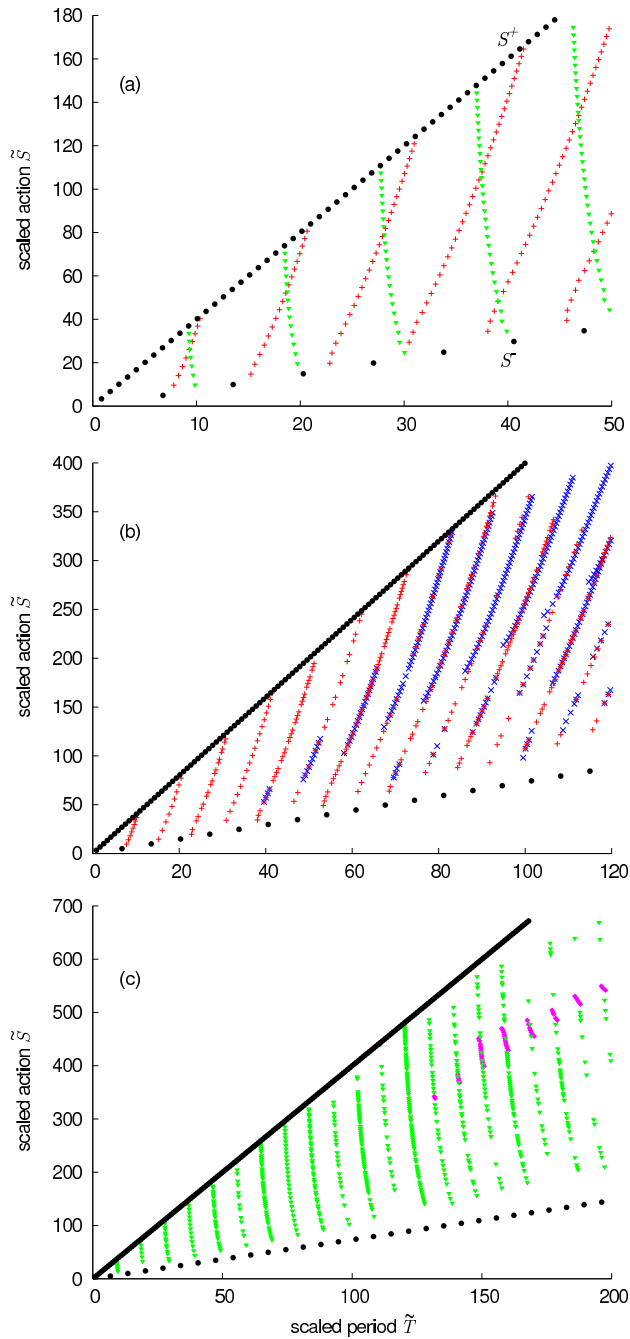}

\caption{Periodic orbits at $\tilde{E}=-1.4$ and $\tilde F=0.5$. The FPOs $S^{\pm}$ and their
repetitions are shown with black circles. (a) the 2-torus POs $T_{2}^{\text{p}}$
(red plus symbols) and $T_{2}^{\text{n}}$ (green triangles). (b)
the $T_{2}^{\text{p}}$ and their 3-torus partners $T_{3}^{\text{p}}$
(blue crosses). (c) the $T_{2}^{\text{n}}$ and their 3-torus partners
$T_{3}^{\text{n}}$ (magenta diamonds).}

\label{fig:TS}
\end{figure}

Our investigation of the POs in the crossed-fields hydrogen atom starts
with a numerical search for periodic orbits. This search, which is
described in detail in Appendix~\ref{app:search}, produces a long
and unstructured list of POs. Several such lists are available in
the literature (see, e.g., \cite{QuasiLandau,Wangplanar,FloethmannKepler,FloethmannMoon}),
but no comprehensive ordering scheme for POs in three degrees of freedom
has been proposed so far, and \emph{a priori} it is not even clear
if one exists. However, in Fig.~\ref{fig:TS} a clear structure underlying
the family of POs becomes apparent: The periods and actions of most
POs fall into well-separated series. (A few orbits that arise from
secondary bifurcations and do not fit this pattern were omitted from
the figure.) In a recent publication \cite{Gekle1} we argued that
the POs arise from the breakup of invariant tori (i.e., they are $N$-torus
POs in the terminology of Sec.~\ref{sec:review}) and that the series
structure provides evidence for a hierarchical ordering of POs that
reflects the hierarchy of invariant tori in an integrable limit of
the dynamics. In the following, we will briefly summarize the conclusions
reached in \cite{Gekle1}, describe in detail the computational procedures
that justify them and demonstrate how the ordering of POs can be used
to gain insight into the topology of higher-dimensional phase space
structures. In the course of the exposition, the nomenclature used
in Fig.~\ref{fig:TS} will be made clear.

\begin{figure}
\begin{center}\includegraphics[width=1\columnwidth]{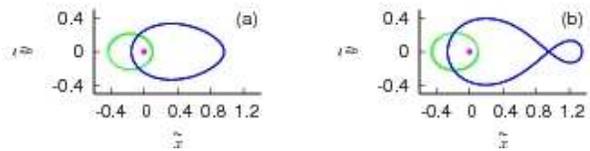}\end{center}

\caption{The two planar FPOs $S^{+}$ (green) and $S^{-}$ (blue) at $\tilde{E}=-1.5$
(a) and $\tilde{E}=-1.4$ (b), where $S^{-}$ is already strongly
deformed. The position of the nucleus is marked by a dot.}
\label{fig:Kepler}
\end{figure}

The three shortest or fundamental periodic orbits (FPOs) of the crossed-fields
hydrogen atom were found in \cite{FloethmannKepler,FloethmannMoon}.
They do not arise from the breakup of a higher-dimensional torus and
therefore represent the 1-torus POs in the hierarchy. Two of these,
labeled $S^{+}$ and $S^{-}$ in \cite{FloethmannMoon}, serve as
organizing centers for the entire hierarchy. Both of them are planar.
They are depicted in Fig.~\ref{fig:Kepler}. For very low energies,
$S^{+}$ and $S^{-}$ are shaped like Keplerian ellipses \cite{FloethmannKepler}.
At the energy levels chosen in this work, $S^{+}$ maintains this
form, whereas $S^{-}$ is strongly deformed as shown in Fig.~\ref{fig:Kepler}.
In addition, above the ionization threshold the FPO $S^{-}$,
although still stable, is surrounded by a phase space region of ionizing trajectories
\cite{Turgayionization,TurgayPODS,Janionization}.

\begin{figure}
\includegraphics[width=1\columnwidth,keepaspectratio]{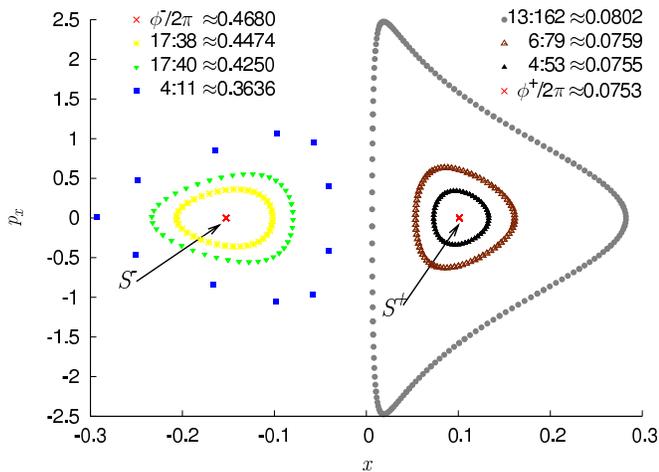}

\caption{\label{fig:SOS}Periodic orbits in the Poincar\'e surface of section
$\tilde{y}=0$ for the planar subsystem, $\tilde{E}=-1.5$, $\tilde{F}=0.5$.
Non-fundamental POs are labeled by their winding ratios.}
\end{figure}

The mechanism that enables the FPOs to organize the hierarchy of $N$-torus
POs around them can best be illustrated if the discussion is temporarily
restricted to the $\tilde{x}$-$\tilde{y}$ plane perpendicular to
the magnetic field. This plane forms an invariant two-dimensional
subsystem. Its dynamics is therefore accessible to an investigation
using a Poincar\'e surface of section such as Fig.~\ref{fig:SOS}.
The plot shows the two FPOs $S^{+}$ and $S^{-}$, each at the center
of an island of stability. They are surrounded by POs of larger periods.
In an obvious way, the latter can be thought of as arising from the
breakup of resonant invariant tori and being interleaved with surviving
KAM tori as described in the previous section. Therefore, the POs
can be labeled with winding numbers and action variables that stem
from the original family of invariant tori, and thus provide access
also to the distribution of the remaining invariant tori, which are
much harder to compute.

An integrable Hamiltonian system with $f>2$ degrees of freedom possesses
an entire hierarchy of invariant tori of dimensions up to $f$. In
every dimension the fully resonant tori break up into isolated POs.
A generic higher-dimensional Hamiltonian system will therefore exhibit
a hierarchy of $N$-torus POs, where at each level the family of $N$-torus
POs is organized by the underlying family of $(N-1)$-torus POs. This
scenario is entirely analogous to the simple two-dimensional example
of Fig.~\ref{fig:SOS}, but beyond two degrees of freedom we cannot
rely on Poincar\'e surface of section plots to diagnose the situation.

In Fig.~\ref{fig:TS} all series end at lines that are formed by
the FPOs and their repetitions. This observation gives us a first
hint that the FPOs indeed serve as organizing centers for the families
of longer POs even in the full, three-dimensional dynamics. This hint,
however, is weak, and it is desirable to characterize the relation
between the FPOs and the families they organize in a more detailed way.

In Ref.~\cite{Gekle1}, we proposed three quantitative criteria to
this end: (i) The stability angles $\phi_{1}^{\pm}$ of the FPOs $S^{\pm}$
(i.e.~the phase angles of the unimodular eigenvalues of their stability
matrices) describe the rotation that each FPO imposes upon its neighborhood.
The winding ratios of the 2-torus POs converge toward $\phi_{1}^{\pm}/2\pi$
as the FPOs are approached. (ii) In the same limit, the action variable
corresponding to the degree of freedom along the FPO converges to
the action of the FPO, and (iii) the action variable for the motion
transverse to the FPO, which is given by the area the (original) invariant
torus encloses in the Poincaré plane, tends to zero. We will use these
criteria, which are derived in a two-dimensional setting, more generally
to characterize the relationship between the $N$-torus POs at different
levels of the hierarchy.

\begin{figure}
\includegraphics[width=1\columnwidth]{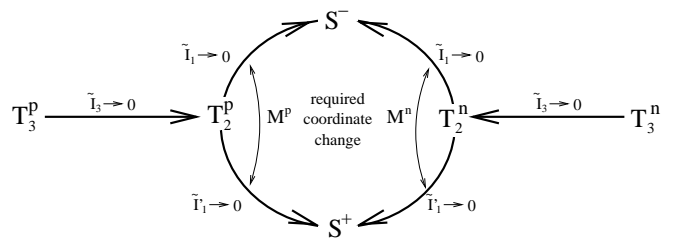}

\caption{\label{fig:genealogy}
The hierarchy of $N$-torus POs in the crossed-fields
hydrogen atom at $\tilde{E}=-1.5$, $\tilde F=0.5$ (and at $\tilde{E}=-1.4$, $\tilde F=0.5$,
except for the collapse on $S^{-}$). The two families of 3-torus
POs $T_{3}^{\text{p}}$ and $T_{3}^{\text{n}}$ collapse onto their
2-torus partners $T_{2}^{\text{p}}$ and $T_{2}^{\text{n}}$ as $\tilde{I}_{3}\rightarrow0$.
The $T_{2}$ in turn collapse onto the FPOs $S^{+}$ and $S^{-}$.
For a correct description of the collapse the appropriate coordinate
system must be used: $\tilde{I}_{1}\rightarrow0$ for $S^{-}$ and
$\tilde{I}_{1}'\rightarrow0$ for $S^{+}$. The coordinate change
is described by the topological invariants $M^{\text{p}}$ and $M^{\text{n}}$.}
\end{figure}

The entire hierarchy of POs that will be reconstructed with the help
of these criteria is illustrated in Fig.~\ref{fig:genealogy}. We
distinguish between two families $T^{\text{p}}$ and $T^{\text{n}}$
of $N$-torus POs that appear as series with positive (p) or negative (n)
slope, respectively, in Fig.~\ref{fig:TS}. Each of these contains
a family $T_{2}^{\text{p,n}}$ of 2-torus POs and a family $T_{3}^{\text{p,n}}$
of 3-torus POs. Since POs which are remnants of the same torus have
nearly identical scaled actions $\tilde{S}$ and periods $\tilde{T}$, each point in Fig.~\ref{fig:TS}
represents one specific broken torus. The FPOs $S^{+}$ and $S^{-}$serve
as organizing centers for both families $T_{2}^{\text{p}}$ and $T_{2}^{\text{n}}$
of 2-torus POs. The POs located in the $\tilde{x}\text{-}\tilde{y}$
symmetry plane (and shown in Fig.~\ref{fig:SOS}) form the family
$T_{2}^{\text{p}}$. The 2-torus POs themselves are found to be limiting
cases of two families $T_{3}^{\text{p}}$ and $T_{3}^{\text{n}}$
of 3-torus POs.

To justify this description of the hierarchy, it is necessary to assign
winding numbers to individual POs and calculate the corresponding
action variables, so that the three criteria can be applied. These
calculations will be presented in Sec.~\ref{sec:WN} and~\ref{sec:tori}.
It will also be shown that the freedom of choice in the definition
of action-angle coordinate systems that was explained in Sec.~\ref{sec:review}
and that at first sight seems to impede the application of the criteria
can in fact be used to assign winding numbers to each torus in the
coordinate system that is best adapted to the local dynamics. The
way how different local coordinate systems transform into one another
then contains global information about the topology of the families
of invariant tori that is represented by the topological invariants
$M^{\text{p}}$ and $M^{\text{n}}$ in Fig.~\ref{fig:genealogy}.

\section{Assignment of winding numbers}
\label{sec:WN}

\subsection{The 2-torus POs $T_{2}^{\text{p}}$ and $T_{2}^{\text{n}}$}
\label{ssec:T2WN}

Winding numbers for the 2-torus POs $T_{2}^{\text{p}}$ and $T_{2}^{\text{n}}$
can be assigned as follows: counting the series in Fig.~\ref{fig:TS}
(a) yields the first winding number $w_{1}$. Counting the POs within
one series from bottom to top yields the second winding number $w_{2}$.
The first series must be assigned $w_{1}=1$ because the repetitions
of the orbits in the leftmost series occur in all higher series and
the $k$-fold repetition of a PO results in another PO with winding
numbers multiplied by $k$ compared to the primitive PO. 

\begin{figure}[b]
\begin{center}\includegraphics[ width=0.7\columnwidth]{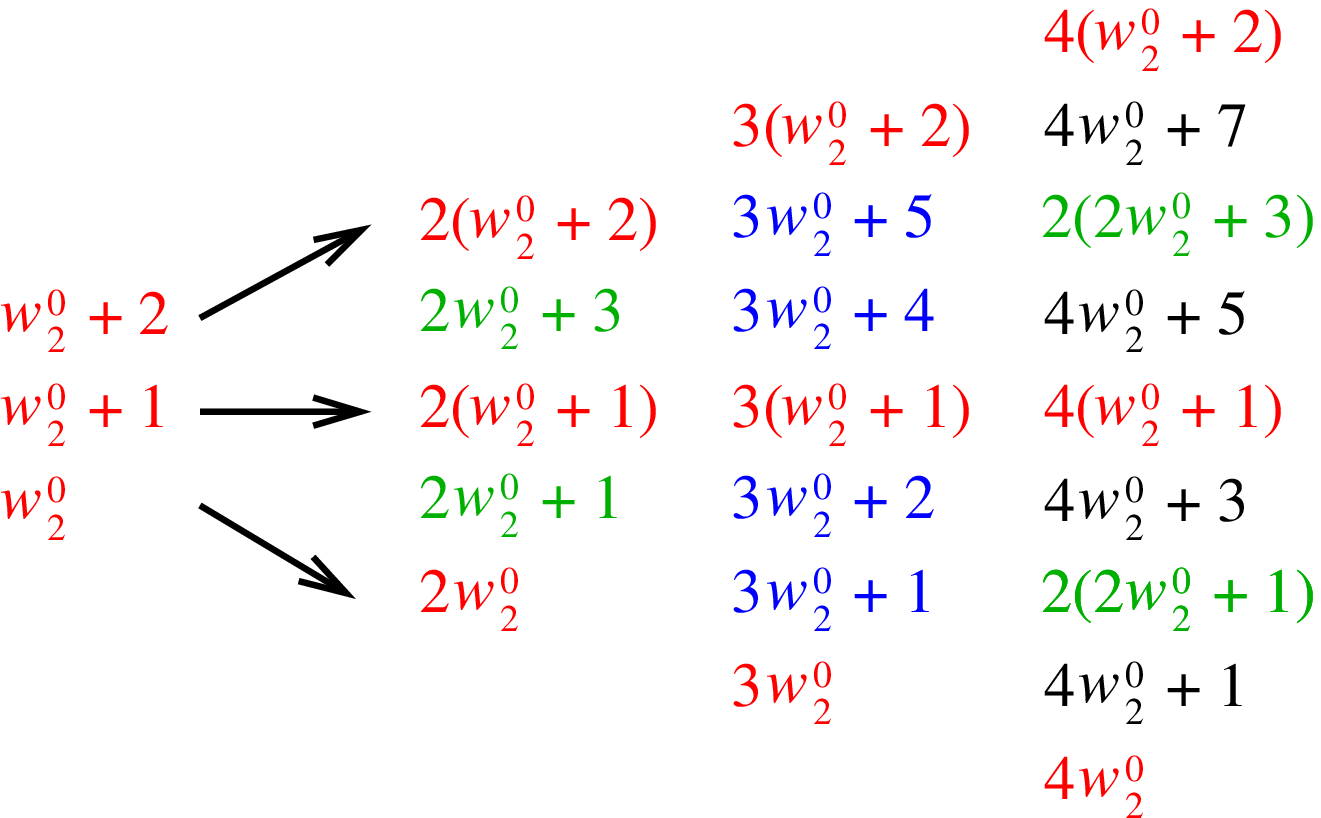}
\end{center}

\caption{The pattern of primitive and repeated POs for the first four series
of 2-torus POs derived theoretically. The first series $w_{1}=1$
contains only primitive POs. The second series $w_{1}=2$ shows an
alternating sequence of primitive POs (odd $w_{2}$) and repetitions
of the POs in the first series (even $w_{2}$). (It cannot be derived
from the multiplication principle if there is an additional primitive
PO at the lower or upper end of the second series.) In the third series
there are blocks of two primitive POs ($w_{2}$ indivisible by 3)
separated by 3-fold repetitions of the first series POs ($w_{2}$
divisible by 3). The fourth series contains primitive POs for odd
values of $w_{2}$, 4-fold repetitions of the first series for positions
with $w_{2}$ divisible by 4 and 2-fold repetitions of the second
series for $w_{2}$ divisible by 2 but not by 4. The scheme can be
extended to arbitrarily large values of $w_{1}$.}

\label{fig:patternrepetitions}
\end{figure}

\begin{figure}[b]
\begin{center}\includegraphics[width=.5\textwidth]{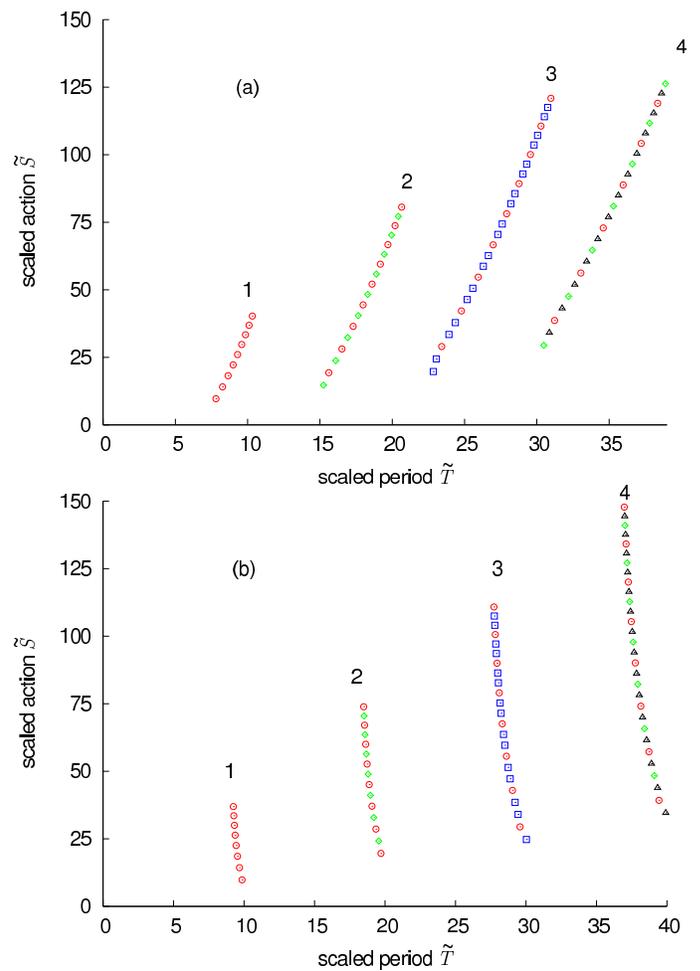}\end{center}

\caption{The pattern of primitive and repeated POs for the first four series
of the $T_{2}^{\text{p}}$ (a) and the $T_{2}^{\text{n}}$ (b). POs
in the first series and their repetitions are shown with red circles,
POs in the second series and their repetitions with green diamonds
and POs from the third and fourth series in blue squares and black
triangles, respectively. The sequence of primitive and repeated POs within the individual series should be compared to the pattern shown in Fig.~\ref{fig:patternrepetitions}, which was derived from the multiplication principle. The numerical results agree perfectly with the theoretical predictions.
}

\label{fig:TSrepcompareE14}
\end{figure}

Using the same multiplication principle, we can perform a first consistency
check on this numbering scheme because it allows us to derive a pattern
for the values of $w_{2}$ across different series. All POs in the
first series are obviously primitive, whereas in higher series primitive
and repeated POs are interleaved. As illustrated in Fig.~\ref{fig:patternrepetitions},
the winding numbers of all orbits in the higher series can be deduced
from those of the first series. The numerical results shown in Fig.~\ref{fig:TSrepcompareE14}
confirm this pattern. Only the base value $w_{2}^{0}$ to be assigned
to the PO with the lowest action in the first series remains undetermined
from these considerations. It will turn out that $w_2^0$ cannot be determined uniquely. Instead, different subsets of POs suggest different values of the winding number $w_2$. As will be explained below, this non-uniqueness reflects the freedom of choice of different action-angle coordinate systems and allows one to assign winding numbers to each torus in the coordinate system that is best adapted to the local dynamics.

\begin{figure}
\begin{center}\includegraphics[width=1\columnwidth]{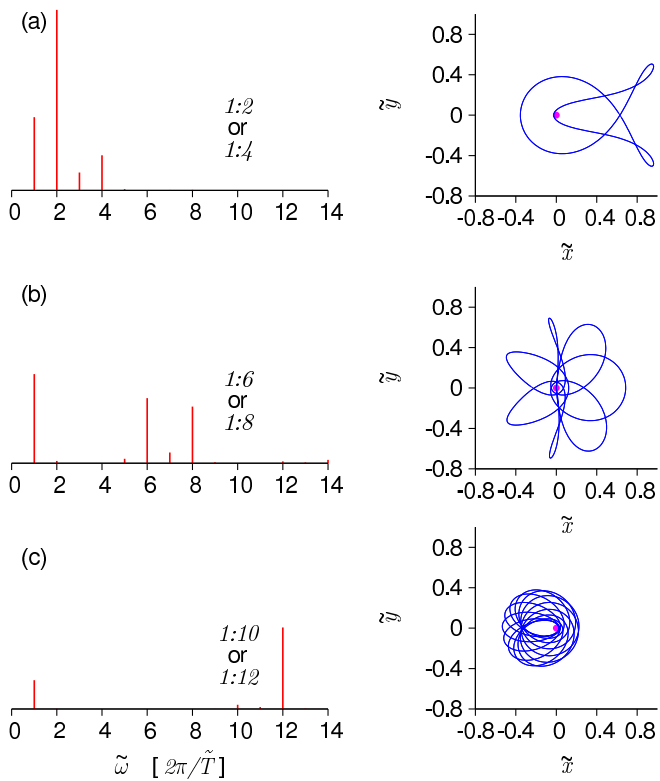}
\end{center}

\caption{Fourier spectra (absolute values of Fourier coefficients) of $\tilde{x}(\tilde{t})$
for some POs of the $T_{2}^{\text{p}}$ family. Orbits are labeled
by winding numbers calculated in two different coordinate systems
(see text). The right column shows the orbits in configuration space
and the position of the nucleus.\label{fig:fourierT2p}}
\end{figure}

To determine $w_{2}^{0}$ we use a Fourier series expansion of the
time series of the Cartesian coordinates, with the frequencies $\tilde{\omega}$
measured in units of the fundamental frequency $\tilde{\omega}_{0}=2\pi/\tilde{T}$.
For a PO with period $\tilde{T}$, these Fourier spectra do not show
peaks at all multiples of the fundamental frequency, but only at those
that correspond to the winding numbers and some of their integer linear
combinations \cite{Fourier1}. For the planar $T_{2}^{\text{p}}$
family it suffices to consider $\tilde{x}(\tilde{t})$, since the
spectra for $\tilde{y}(\tilde{t})$ show peaks at identical positions.
The first peak at $\tilde{\omega}=1\cdot\tilde{\omega}_{0}$ in the
example of Fig.~\ref{fig:fourierT2p} corresponds to the series number
$w_{1}$ of the POs. It is found in the same position for any PO in
a given series. The second major peak corresponds to the second winding
number $w_{2}$. Fig.~\ref{fig:fourierT2p}(a) shows the spectrum
for the PO with the lowest action in the first series, where this
second peak is located at $\tilde{\omega}=2\cdot\tilde{\omega}_{0}$. 

We can therefore set $w_{2}^{0}=2$, which according to Fig.~\ref{fig:patternrepetitions}
determines the values of $w_{2}$ for all orbits. However, as Fig.~\ref{fig:fourierT2p}(b)
and (c) show, the peak corresponding to this value of $w_{2}$ is
not always dominant and can indeed be very weak if $w_{2}$ is large.
Instead, a strong peak arises at a different position and suggests
choosing a different set of winding numbers that has $w_{2}^{0\prime}=4$
and is connected to the previous system by the transformation\begin{eqnarray}
w_{1}^{\prime} & = & w_{1},\nonumber \\
w_{2}^{\prime} & = & 2w_{1}+w_{2}.\label{eqn:trafos}\end{eqnarray}
Both choices are equally viable. They correspond to different choices
of angle coordinates on the original tori. As the Fourier spectra
illustrate, the {}``unprimed'' coordinate system is well adapted
to the dynamics of orbits with low values of $w_{2}$, whereas the
{}``primed'' coordinates describe the dynamics of orbits with high
$w_{2}$.

In view of Fig.~\ref{fig:TS} (a) it is evident that POs with low
$w_{2}$ lie close to $S^{-}$, while POs with high $w_{2}$ are close
to $S^{+}$. The transition indicates that the dominant modes of the
dynamics change from one end of the series to the other: close to
$S^{-}$, these normal modes are the motion parallel and transverse
to the FPO. As the distance from $S^{-}$ grows, the coupling of the
normal modes increases, and as $S^{+}$ is approached, the normal
modes imposed by that FPO, which are different, gain dominance. In
this way, the two FPOs that bound the $T_{2}^{\text{p}}$ family impose
two different systems of angle coordinates $\boldsymbol{\theta}$
and $\boldsymbol{\theta}'$ on the tori in their neighborhood. The
transformation between the two coordinate systems is a topological
invariant that characterizes the family of 2-torus POs. In our case,
it is given by Eq.~(\ref{eqn:trafos}) and characterized by the matrix
\begin{equation}
M^{\text{p}}=\begin{pmatrix}1 & 0\\
2 & 1\end{pmatrix}.\end{equation}
 As anticipated in Sec.~\ref{sec:review}, $M^{\text{p}}$ is an
integer matrix with unit determinant. 

\begin{figure}
\begin{center}\includegraphics[width=\columnwidth]{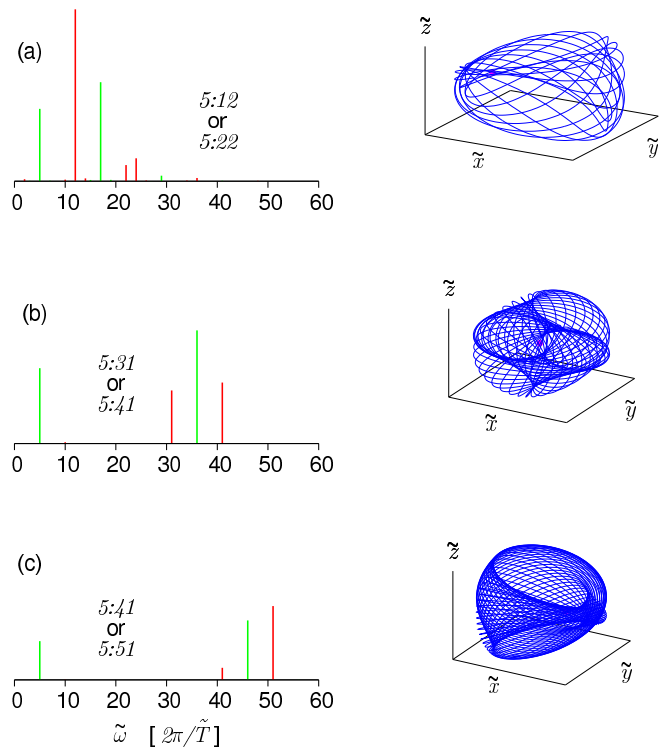}\end{center}

\caption{Fourier expansions of $\tilde{x}(\tilde{t})$ (red) and $\tilde{z}(\tilde{t})$
(green) for POs from the $T_{2}^{\text{n}}$ family. Peaks not associated
with winding numbers are located at integer linear combinations thereof
\cite{Fourier1}.}

\label{fig:fourierT2n}
\end{figure}

Using a similar method, one can assign winding numbers to the POs of the $T_{2}^{\text{n}}$ family. These POs are not restricted to the $x$-$y$ symmetry plane, they are truly three-dimensional in Cartesian coordinate space, so that a Poincar\'e surface of section plot similar to Fig.~\ref{fig:SOS} cannot be obtained. Nevertheless, the Fourier spectra of $\tilde{x}(\tilde{t})$ and $\tilde{z}(\tilde{t})$ provide the information needed for a complete
assignment of winding numbers, as shown in Fig.~\ref{fig:fourierT2n}.
It is again found that the dominant normal modes are different for
POs at the lower end of the series, close to $S^{-}$, and at the
upper end of the series, close to $S^{+}$. The invariant $M^{\text{n}}=M^{\text{p}}$ that
characterizes the transition between the two limits takes the same
value for both families.

\subsection{The 3-torus POs $T_{3}^{\text{p}}$ and $T_{3}^{\text{n}}$}

\begin{figure}
\begin{center}\includegraphics[width=.5\textwidth]{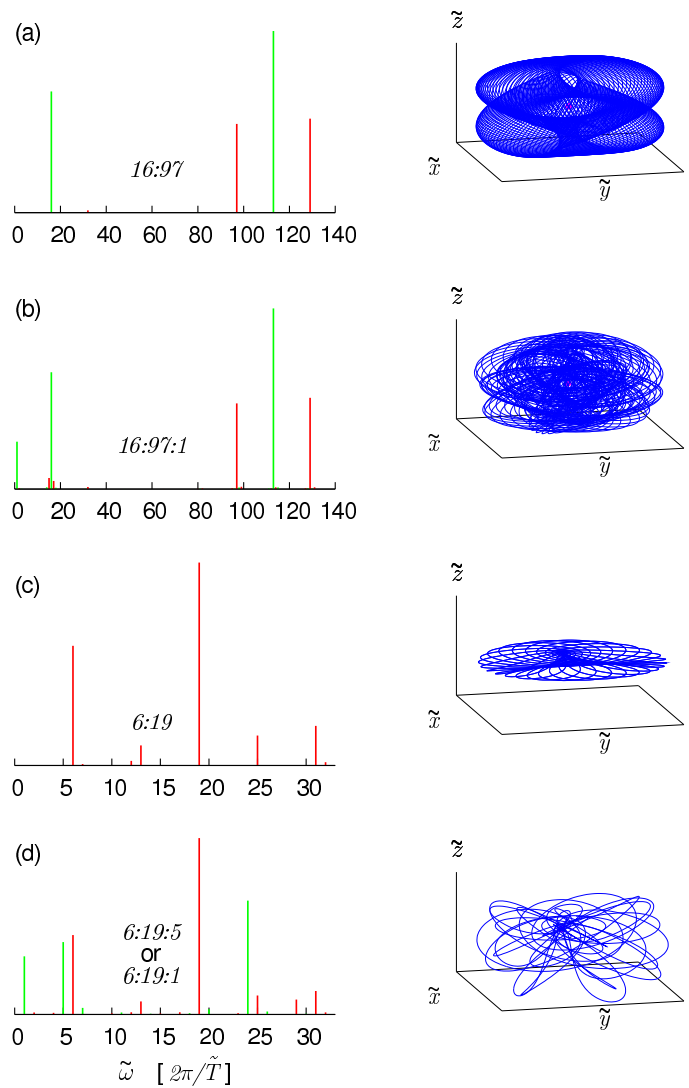}\end{center}

\caption{Fourier spectra of $\tilde{x}(\tilde{t})$ (red) and $\tilde{z}(\tilde{t})$
(green). The second winding number is given in the unprimed coordinate
system $w_{2}^{0}=2$. (a) A 2-torus PO in the $T_{2}^{\text{n}}$
family and (b) its 3-torus partner from $T_{3}^{\text{n}}$. (c) A
2-torus PO from the $T_{2}^{\text{p}}$ family and (d) its 3-torus
partner from $T_{3}^{\text{p}}$.}

\label{fig:fourierT3}
\end{figure}

Apart from the 2-torus POs, Fig.~\ref{fig:TS} (b) and (c) show POs
that are generated in the destruction of 3-tori. They are arranged
in the same series as the 2-torus POs, and each 3-torus PO is intimately
related to a 2-torus ``partner'' with almost identical period and
action. Such partners have identical winding numbers $w_{1}$ and
$w_{2}$, but 3-torus POs possess a third winding number $w_{3}$,
which distinguishes different 3-torus PO partners of the same 2-torus
PO. It manifests itself in an additional peak in the Fourier spectra
and can therefore be assigned by a straightforward extension of the
technique used to classify the 2-torus POs. For the $T_{3}^{\text{n}}$
as in Fig.~\ref{fig:fourierT3} (a) and (b) the situation is unambiguous
since there is only one strong additional peak. The location of this
peak provides the winding number $w_{3}$. For POs of the $T_{3}^{\text{p}}$
family (see Fig.~\ref{fig:fourierT3} (c) and (d)), however, two
additional peaks of comparable magnitude arise, giving us two possibilities
to assign the third winding number. In Fig.~\ref{fig:fourierT3}
(d), the two possibilities are $w_{3}=5$ and $w_{3}'=1$. As for
the 2-torus POs, the two systems of winding numbers correspond to
different angle coordinate systems on the original tori. In our case,
the transformation between the two systems is given by: \begin{eqnarray}
w_{3}' & = & w_{1}-w_{3}.\label{eqn:trafou}\end{eqnarray}

Note that there is no obvious relation between the classification
by winding numbers that becomes possible though the present analysis and the geometrical appearance of the POs in configuration
space.

\subsection{The stability of the POs}

The stability of a PO in a Hamiltonian system with three degrees of freedom is characterized by a $4\times 4$ symplectic stability matrix, which describes the linearized dynamics transverse to the PO \cite{Arnold,MacKay,Meiss}. Because the stability matrix is real and symplectic, with each eigenvalue $\lambda$ its inverse $1/\lambda$ and its complex conjugate $\lambda^\ast$ must also be eigenvalues. Therefore, if the eigenvalues of the stability matrix are different from $\pm 1$, they must belong to either (i) an elliptic pair $e^{\pm i\varphi}$ of complex conjugate eigenvalues with unit modulus, (ii) a hyperbolic pair $\lambda$, $1/\lambda$ of real eigenvalues, or (iii) a loxodromic quartet $\lambda$, $1/\lambda$, $\lambda^\ast$, $1/\lambda^\ast$ of complex eigenvalues. A PO is stable if and only if all eigenvalues of its stability matrix have unit modulus.
 
On an $N$-dimensional resonant torus in an integrable system, POs occur in continuous $N$-parameter families. They therefore have marginal stability, i.e., all four eigenvalues of their stability matrix equal one. In a non-integrable system only a
small number of isolated POs remains. The eigenvalues of their stability
matrix occur in elliptic (e) or hyperbolic (h) pairs. In our numerical studies we did not find
any loxodromic quartets for $N$-torus POs. In some
cases our search algorithm finds a large number of POs all
of which originate from the breakup of the same torus and have almost marginal
stability. In these cases, the splitting of the original torus is
so small that we cannot resolve isolated POs within the given numerical
precision. In the other cases, whenever our numerical PO search allows
us to identify all members in a quadruplet resulting from the breakup
of a 3-torus, we find that it contains all four stability combinations
ee, eh, he and hh, in accordance with \cite{Meiss}. 

\begin{figure}
\begin{center}\includegraphics[width=1\columnwidth]{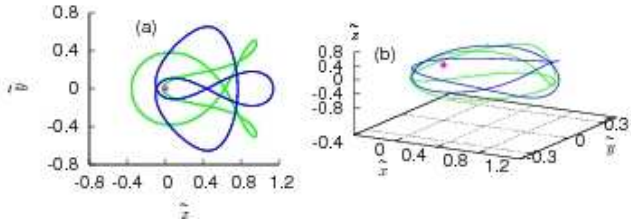}\end{center}

\caption{Elliptic (green) and hyperbolic (blue) POs with winding numbers 1:2
(a) in the $T_{2}^{\text{p}}$ family and (b) in the $T_{2}^{\text{n}}$
family.}

\label{fig:paireh}
\end{figure}

The 2-torus POs occur as doublets, of which Fig.~\ref{fig:paireh}
shows an example. One pair of stability eigenvalues is approximately
the same for both members of a doublet. It corresponds to the motion
transverse to the original torus. With very few exceptions, the dynamics
in this direction is stable. In the direction along the original torus
one partner is elliptic while the other is hyperbolic in accordance
with the situation of the Poincar\'{e}-Birkhoff theorem \cite{PBtheorem}
in two degrees-of-freedom. Figs.~\ref{fig:fourierT2p}, \ref{fig:fourierT2n},
\ref{fig:fourierT3} show elliptic (e) POs for the $T_{2}$ and completely
elliptic (ee) for the $T_{3}$ in those cases where isolated POs can
be identified.

\subsection{Symmetry considerations}

The Hamiltonian (\ref{eqn:Hamiltonian}) possesses three discrete
symmetries, namely the reflection symmetry with respect to the $x$-$y$
plane ($z$-parity), the $y$-parity with additional time reversal
and the combination of both. Due to the $z$-parity, the $x$-$y$ plane
constitutes a two-degree-of-freedom subsystem. Any PO outside this
plane must either be itself invariant under the symmetry transformation
or possess a partner related to it via the symmetry transformation. 

We find the following connection between winding numbers and symmetry:
A PO in the $T_{3}^{\text{p}}$ family is symmetric under the $z$-parity
transformation if and only if its winding numbers $w_{1}$ and $w_{2}$
are even. In the $T_{2}^{\text{n}}$ and $T_{3}^{\text{n}}$ families,
the POs with $w_{1}$ odd and $w_{2}$ even are symmetric under $z$-parity, while all
others are not. For both families of 3-torus POs the third winding
number $w_{3}$ has no influence on the symmetry properties.

\section{Calculation of action variables}

\label{sec:tori}

In this section we compute the scaled action variables $\mathbf{\tilde{I}}$
for each family of POs. 

The total action of a PO can be expressed in terms of the individual
actions $\mathbf{\tilde{I}}$: \begin{equation}
\tilde{S}=\mathbf{w}\cdot\mathbf{\tilde{I}}\left(\mathbf{w}\right).\end{equation}
 For any multiple $\alpha$ of the PO, this becomes: \begin{equation}
\alpha\tilde{S}=\alpha\mathbf{w}\cdot\mathbf{\tilde{I}}\left(\alpha\mathbf{w}\right),\end{equation}
with the same action variables $\tilde{\mathbf{I}}(\alpha\mathbf{w})=\tilde{\mathbf{I}}(\mathbf{w)}$as
before. Thus, the $\mathbf{\tilde{I}}$ for the $T_{3}$ families
can be written as functions of any two frequency ratios (which by
virtue of Eq.~(\ref{eqn:ratios}) equal the ratios of the winding
numbers). Here we choose \begin{equation}
\mathbf{\tilde{I}}\left(w_{1},w_{2},w_{3}\right)=\mathbf{\tilde{I}}\left(\frac{w_{1}}{w_{2}},\frac{w_{3}}{w_{2}}\right).\end{equation}
 To calculate the action variables $\mathbf{\tilde{I}}$ numerically
we select three POs close to each other in the $\left(w_{1}/w_{2},\; w_{3}/w_{2}\right)$
plane, so that the action variables are approximately constant in the small triangular
area in between. Under this assumption, we obtain a system of equations
\begin{equation}
\begin{array}{rll}
\tilde{S}^{(1)} & = & w_{1}^{(1)}\tilde{I}_{1}+w_{2}^{(1)}\tilde{I}_{2}+w_{3}^{(1)}\tilde{I}_{3},\\
\tilde{S}^{(2)} & = & w_{1}^{(2)}\tilde{I}_{1}+w_{2}^{(2)}\tilde{I}_{2}+w_{3}^{(2)}\tilde{I}_{3},\\
\tilde{S}^{(3)} & = & w_{1}^{(3)}\tilde{I}_{1}+w_{2}^{(3)}\tilde{I}_{2}+w_{3}^{(3)}\tilde{I}_{3},\end{array}\label{eqn:3toriactions}
\end{equation}
which can be solved for $\tilde{I}_{1}$, $\tilde{I}_{2}$, and $\tilde{I}_{3}$.
These values are then assigned to be the function values at the barycenter
of the triangle. The process is repeated for all POs in the corresponding
family and the results are used to calculate an interpolation function
using a modified Shepard's method \cite{NAG}. 

\begin{figure}
\begin{center}\includegraphics[width=1\columnwidth,keepaspectratio]{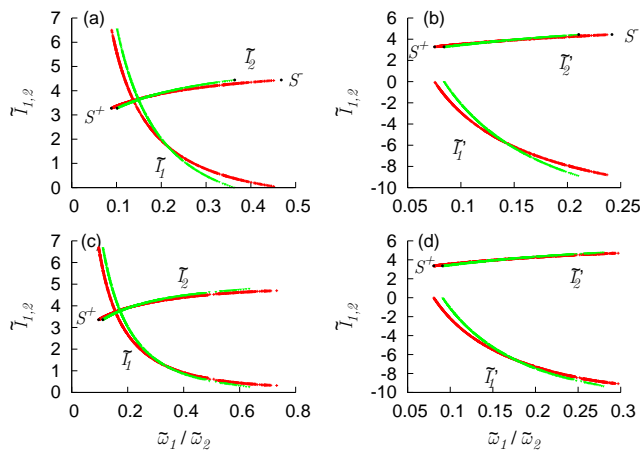}
\end{center}

\caption{The actions $\tilde{I}_{1}$ and $\tilde{I}_{2}$ for the $T_{2}^{\text{p}}$
(red plus symbols) and $T_{2}^{\text{n}}$ (green triangles). (a)
Actions at $\tilde{E}=-1.5$, $\tilde F=0.5$ in the unprimed action-angle coordinate system with $w_2^02$. The winding
numbers are suitable to describe the approach to $S^{-}$, where $\tilde{I}_{1}$
tends to zero. The stability angles of the FPO $S^{+}$ have been
transformed using Eq.~(\ref{eqn:phitransform}). (b) The same data
in the primed coordinate system with $w_2^0=4$. The situation around the
FPOs is reversed. (c) Actions at $\tilde{E}=-1.4$, $\tilde F=0.5$ (unprimed coordinates). The
$T_{2}^\text{p,n}$ do not collapse onto $S^{-}$ due to the ionizing region
around that FPO. (d) Actions at $\tilde{E}=-1.4$, $\tilde F=0.5$ (primed coordinates).}

\label{fig:2toriactions}
\end{figure}

For the 2-torus POs we follow a similar procedure, except that we
need to consider only a single winding ratio $w_{1}/w_{2}$. At the
energy $\tilde{E}=-1.5$, we thus obtain the results shown in Fig.~\ref{fig:2toriactions}
(a), which also displays the actions and stability angles $\phi_{1,2}$
of the FPOs. The limiting values for high frequency ratios of the
$T_{2}^{\text{p}}$ and $T_{2}^{\text{n}}$ families coincide with
$\phi_{1}/2\pi$ and $\phi_{2}/2\pi$ of $S^{-}$, respectively. At
the same time, the action variable $\tilde{I}_{2}$ converges toward
the action of the FPO $S^{-}$, whereas the action variable $\tilde{I}_{1}$
vanishes. According to the three criteria listed in Sec.~\ref{sec:orga},
we can thus conclude that $S^{-}$ serves as an organizing center
for both families $T_{2}^{\text{p}}$ and $T_{2}^{\text{n}}$ and
that the action variable $\tilde{I}_{2}$ and $\tilde{I}_{1}$ correspond
to the normal modes along and transverse to $S^{-}$. (Notice that
in three degrees of freedom a stable FPO has two pairs of unimodular
stability eigenvalues, so that the collapse scenario of Fig.~\ref{fig:SOS}
can take place in two transverse degrees of freedom independently,
giving rise to two families of 2-torus POs.) For the planar family
$T_{2}^{\text{p}}$, this result merely confirms the conclusion that
we could already draw from the Poincar\'e surface of section plot in
Fig.~\ref{fig:SOS}. By contrast, the family $T_{2}^{\text{n}}$,
although it consists of 2-torus POs, is not contained in any two-dimensional
subsystem that could be described without an intimate study of the dynamics.
Its analysis is therefore beyond the reach of a Poincar\'e
plot. Nevertheless, \ref{fig:2toriactions}(a) demonstrates that the
relations of the families $T_{2}^{\text{p}}$ and $T_{2}^{\text{n}}$
to the FPO $S^{-}$ are entirely analogous and that $S^{-}$ organizes
$T_{2}^{\text{n}}$ just as much as $T_{2}^{\text{p}}$.

As described in Sec.~\ref{ssec:T2WN}, the winding numbers $\mathbf{w}$
used in Fig.~\ref{fig:2toriactions} (a) are ill-suited to describe
the approach to $S^{+}$, and the winding numbers $\mathbf{w'}$ given
by~(\ref{eqn:trafos}) should be used instead. Indeed, none of the
actions $\tilde{I}_{1}$ and $\tilde{I}_{2}$ tends to zero in the
limit of low frequency ratios. Nevertheless, $S^{+}$ can be located
in Fig.~\ref{fig:2toriactions} (a) if its stability angles $\phi'$,
which naturally arise in the $\mathbf{w}'$ system, are transformed
to the $\mathbf{w}$ system just as if they were regular frequency
ratios by applying the inverse transformation of (\ref{eqn:trafos}):
\begin{equation}
\frac{w_{1}}{w_{2}}=\frac{w_{1}'}{w_{2}'-2w_{1}'}=\frac{\frac{w_{1}'}{w_{2}'}}{1-2\frac{w_{1}'}{w_{2}'}}\end{equation}
 and therefore \begin{equation}
\frac{\phi}{2\pi}=\frac{\frac{\phi'}{2\pi}}{1-2\frac{\phi'}{2\pi}}.\label{eqn:phitransform}\end{equation}

The transformed stability angles $\phi$ in Fig.~\ref{fig:2toriactions}
(a) demonstrate that the $T_2^\text{p,n}$ emanate from $S^{+}$ just as they
emanate from $S^{-}$ in the limit of high winding ratios. This result
can be illustrated more clearly if the actions are calculated in the
$\mathbf{w'}$ coordinate system, as shown in Fig.~\ref{fig:2toriactions}(b).
Here, the situation around the FPOs reverses: $\tilde{I}_{1}'$ tends
to zero as $S^{+}$ is approached and remains finite in the vicinity
of $S^{-}$. (Note that by virtue of Eq.~(\ref{eqn:actionstrafo})
there can exist coordinate systems in which the action variables take
negative values.) In Fig.~\ref{fig:2toriactions}(b), the true stability
angles of $S^{+}$ can be used to indicate the lower limits of the
winding ratios, whereas the stability angles of $S^{-}$ need to be
transformed with the inverse of Eq.~(\ref{eqn:phitransform}). 

This symmetry demonstrates that locally the collapse of the $T_2^\text{p,n}$
onto $S^{+}$ and $S^{-}$ looks the same when described in suitable
local coordinates. The nontrivial topology of the families $T_{2}^{\text{p}}$
and $T_{2}^{\text{n}}$ that is represented by the invariants $M^{\text{p,n}}$
becomes visible only if the entire families, including both limiting
FPOs, are studied, cf.~Fig.~\ref{fig:genealogy}. 

Fig.~\ref{fig:2toriactions} (c) and (d) shows the situation at
$\tilde{E}=-1.4$ and $\tilde F=0.5$, above the ionization saddle point. Here $S^{-}$
is surrounded by an area filled with ionizing trajectories \cite{Turgayionization,TurgayPODS,Janionization}.
This ionizing region prevents the $T_{2}^\text{p,n}$ from collapsing onto $S^{-}$.
Yet, they still emanate from $S^{+}$. 

\begin{figure}
\includegraphics[width=.5\textwidth]{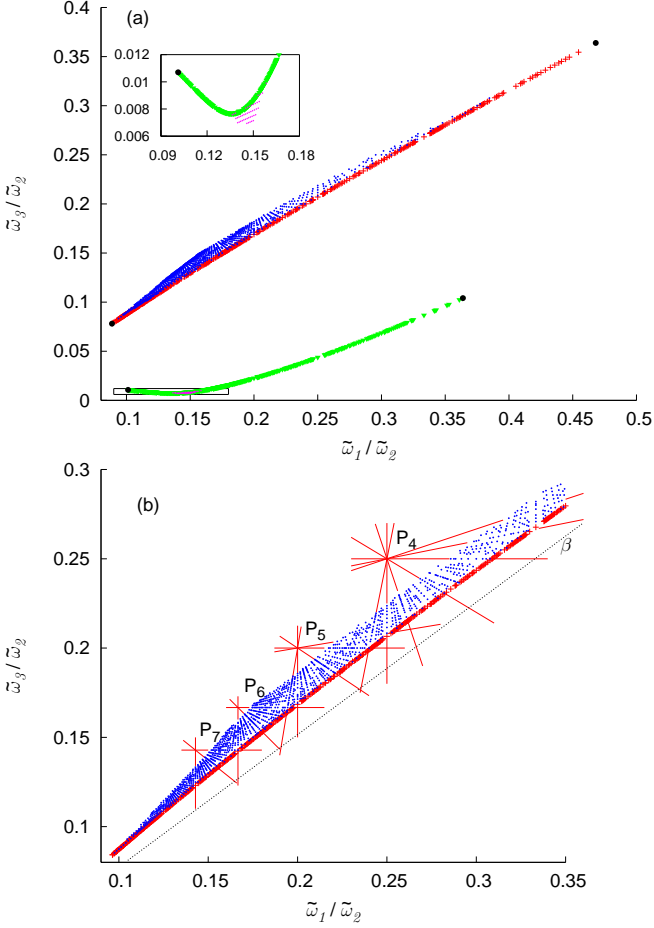}

\caption{(a) Frequency map for $N$-torus POs at $\tilde{E}=-1.5$, $\tilde F=0.5$. The 3-torus
POs $T_{3}^{\text{p}}$ (blue crosses) and $T_{3}^{\text{n}}$ (magenta
diamonds) are bounded by the corresponding 2-torus POs $T_{2}^{\text{p}}$
(red plus symbols) and $T_{2}^{\text{n}}$ (green triangles). The
2-torus POs in turn are bounded by the FPOs $S^{\pm}$ (black dots).
(b) The $T_{3}^{\text{p}}$ at $\tilde{E}=-1.4$, $\tilde F=0.5$ with some of the
most prominent resonance lines highlighted. They intersect in the
points $P_{i}=\left(\frac{1}{i},\frac{1}{i}\right)$.}

\label{fig:stability}
\end{figure}

Fig.~\ref{fig:stability} (a) shows the frequency ratios of 2-
and 3-torus POs. For the 2-torus POs, the missing ratio $\tilde{\omega}_{3}/\tilde{\omega}_{2}$
is replaced with the stability angle that describes the dynamics transverse
to the original torus, normalized by $2\pi w_{2}$. For the FPOs this
angle is given by $\phi_{1}-\phi_{2}$. The surroundings of a long
stable PO can rotate multiple times around the PO during one period.
Stability angles, however, can be calculated only modulo $2\pi$.
Therefore an integer number of $2\pi$ needs to be added to the stability
angles of long POs. These stability angles arise in the $w_{3}'$
coordinate system and thus need to be transformed with the inverse
of Eq.~(\ref{eqn:trafou}). From Fig.~\ref{fig:stability} (a) it
becomes apparent that the $T_{3}$ approach the $T_{2}$ in a suitable
limit just as the $T_{2}$ approach the FPOs. 

Frequency maps such as Fig.~\ref{fig:stability}(b) have been found
in, e.g., \cite{FMA1,Resonance1} to capture the essential dynamics
of a multidimensional system. Their most prominent features are the
resonance lines, each of which is given by a resonance condition of
the form
\begin{equation}
  \sum_{i=1}^f m_{i}\tilde{\omega}_{i}=0
 \label{eqn:reslines}
 \end{equation}
 with integer coefficients $m_{i}$. Their significance arises from
the fact they they represent cantori that act as partial barriers
to phase space transport. Diffusive trajectories such as those used
in \cite{FMA1,Resonance1} will stick to the cantori for a long time
before leaving their neighborhood and thereby cause the prominence
of the corresponding resonance lines in the frequency map. 

The periodic orbits shown in Fig.~\ref{fig:stability} are the remnants
of fully resonant tori, on which $f-1$ independent resonance conditions
of the form (\ref{eqn:reslines}) are satisfied. (They can be derived
from the condition (\ref{eqn:ratios})). 3-torus POs, therefore, will
be located at the intersections of two resonance lines in the frequency
map. Since POs are non-wandering, they belong to the rigid background
through which the diffusion of generic trajectories takes place \cite{Arnoldweb}.
As Fig.~\ref{fig:stability} clearly shows, the set of POs presents
the web of dominant resonances just as clearly as the set of diffusive
trajectories that is customarily used.

\begin{figure}
\includegraphics[width=1\columnwidth]{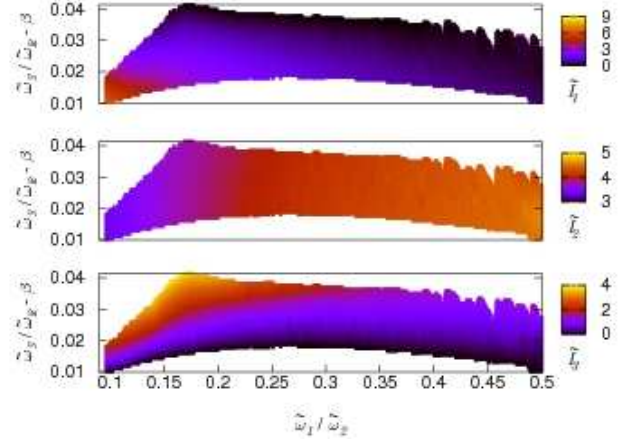}

\caption{\label{fig:3toriactions}Action variables for the $T_{3}^{\text{p}}$
family at $\tilde{E}=-1.4$, $\tilde F=0.5$. At the lower boundary, where the $T_{3}^{\text{p}}$
are bounded by the $T_{2}^{\text{p}}$, $\tilde{I}_{3}$ tends to
zero while $\tilde{I}_{1,2}$ tend toward the values obtained from
the 2-torus POs. For clarity the dotted line $\beta$ from Fig.~\ref{fig:stability}
(b) has been subtracted on the vertical axis.}
\end{figure}

The three action variables that characterize the 3-torus POs are displayed
in Fig.~\ref{fig:3toriactions} for the $T_{3}^{\text{p}}$ family
as functions of the two independent frequency ratios $\tilde{\omega}_{1}/\tilde{\omega}_{2}$
and $\tilde{\omega}_{3}/\tilde{\omega}_{2}$. They were calculated
from Eq.~(\ref{eqn:3toriactions}) as described above. The action
variables $\tilde{I}_{1}$ and $\tilde{I}_{2}$ obtained from the
3-torus POs converge towards those found for the 2-torus POs, while
$\tilde{I}_{3}$ tends to zero as the lower boundary is approached.
We thus find all three criteria of Sec.~\ref{sec:orga} satisfied
in this higher-dimensional situation: The tranverse frequency ratio
$\tilde{\omega}_{3}/\tilde{\omega}_{2}$ of the $T_{3}^{\text{p}}$
tends to the transverse stability angle of the $T_{2}^{\text{p}}$,
the longitudinal action variables $\tilde{I}_{1}$ and $\tilde{I}_{2}$
of the 3-tori approach those of the 2-tori, and the transverse action
variable $\tilde{I}_{3}$ vanishes. We can therefore conclude that
the $T_{2}^{\text{p}}$ family serves as an organizing center for
the $T_{3}^{\text{p}}$, and we can identify the degree of freedom
corresponding to $\tilde{I}_{3}$ as being transverse to the family
of 2-tori. Similar results can be obtained for the collapse of the
$T_{3}^{\text{n}}$ family onto $T_{2}^{\text{n}}$, thus giving a
full justification for all relations depicted in Fig.~\ref{fig:genealogy}.

Because of these relations, additional data points at the lower boundary
in Fig.~\ref{fig:3toriactions} can be obtained from the $T_{2}^{\text{p}}$
(and are included in the figure). For these, the missing frequency
ratio $\tilde{\omega}_{3}/\tilde{\omega}_{2}$ is replaced with the
stability angle according to Fig.~\ref{fig:stability}, the action
variables $\tilde{I}_{1}$ and $\tilde{I}_{2}$ are taken from Fig.~\ref{fig:2toriactions},
and $\tilde{I}_{3}$ is zero.

\section{Torus quantization}
\label{sec:quant}

The action variables calculated in the preceding paragraph provide
the basis for an Einstein-Brillouin-Keller (EBK) torus quantization
of the hydrogen atom in crossed electric and magnetic fields. The
way of recording a quantum spectrum that is best suited to semiclassical
investigations of atomic spectra is scaled-energy spectroscopy, which
has for that reason been used in most experimental and theoretical
work \cite{QuasiLandau,ExactQMTaylor,Experimentsymmetrybreak}. It
offers the advantage that the underlying classical dynamics does not
change across the spectrum. A scaled spectrum consists of those values
of the scaling parameter $B^{-1/3}$ that characterize the quantum
states for given scaled energy $\tilde{E}$ and scaled electric field
strength $\tilde{F}$. 

The phase space volume filled by 3-torus POs from the $T_{3}^{\text{p}}$
family is considerably larger than the volume filled by $T_{3}^{\text{n}}$,
as can be seen by the number of respective POs for example in Fig.~\ref{fig:TS}.
Thus, the largest part of the spectrum is also obtained from the quantization
of the actions belonging to $T_{3}^{\text{p}}$. 

The EBK quantization condition reads \cite{Einstein,Brillouin1,Keller1,Percival77,Brack}:
\begin{equation}
I_{i}=\oint_{\gamma_{i}}\mathbf{p}d\mathbf{q}=2\pi\hbar\left(n_{i}+\frac{\alpha_{i}}{4}\right)\;\;,\qquad i=1,\ldots,f,\end{equation}
 where the $\gamma_i$ are the fundamental loops on the torus used in the definition~(\ref{eqn:action}) of the action variables, $n_{i}$ are integer quantum numbers and $\alpha_{i}$ the
corresponding Maslov indices. As shown above, one can obtain the action integrals from PO data without having to determine the paths $\gamma_i$. In our case, we calculate
the scaled action variables $\tilde{I}_{i}$ which need
to be rescaled to obtain the true actions $I_{i}=\tilde{I}_{i}B^{-1/3}$.
Thus, the quantization condition reads, with $\hbar=1$ in atomic
units: \begin{equation}
\tilde{I}_{i}\left(\frac{w_{1}}{w_{2}},\frac{w_{3}}{w_{2}}\right)B^{-1/3}=2\pi\left(n_{i}+\frac{\alpha_{i}}{4}\right).\label{eqn:EBK}\end{equation}
 We derive the Maslov indices $\alpha_{i}$ from two simple considerations:
In the limit of small $\tilde{I}_{1}$ and $\tilde{I}_{3}$, the action variable $\tilde{I}_{2}$ and its conjugate angle
correspond to the motion along the elliptic FPOs, which clearly has rotational
character and requires $\alpha_{2}=0$. The other two modes are transverse
to this fundamental motion and are thus expected to have vibrational
character, which leads to $\alpha_{1}=\alpha_{3}=2$. 

Inverting Eq.~(\ref{eqn:EBK}) to calculate $B^{1/3}$ gives: \begin{equation}
B^{-1/3}=\frac{2\pi\left(n_{i}+\frac{\alpha_{i}}{4}\right)}{\tilde{I}_{i}\left(\frac{w_{1}}{w_{2}},\frac{w_{3}}{w_{2}}\right)}.\label{eqn:scalParam}\end{equation}
A quantization condition for the scaled spectrum is obtained by observing
that $B^{-1/3}$ must take the same value for all three degrees of
freedom. We can thus calculate the frequency ratios corresponding
to the state with given quantum numbers $n_{1},\: n_{2},\: n_{3}$
from the following set of equations

\begin{eqnarray}
\frac{\tilde{I}_{1}\left(\frac{w_{1}}{w_{2}},\frac{w_{3}}{w_{2}}\right)}{2\pi\left(n_{1}+\frac{\alpha_{1}}{4}\right)}-\frac{\tilde{I}_{2}\left(\frac{w_{1}}{w_{2}},\frac{w_{3}}{w_{2}}\right)}{2\pi\left(n_{2}+\frac{\alpha_{2}}{4}\right)} & = & 0,\label{eqn:system1}\\
\frac{\tilde{I}_{1}\left(\frac{w_{1}}{w_{2}},\frac{w_{3}}{w_{2}}\right)}{2\pi\left(n_{1}+\frac{\alpha_{1}}{4}\right)}-\frac{\tilde{I}_{3}\left(\frac{w_{1}}{w_{2}},\frac{w_{3}}{w_{2}}\right)}{2\pi\left(n_{3}+\frac{\alpha_{3}}{4}\right)} & = & 0\label{eqn:system2}\end{eqnarray}
and then compute the value of the scaling parameter from~(\ref{eqn:scalParam}).

\begin{figure*}
\includegraphics[width=\textwidth,keepaspectratio]{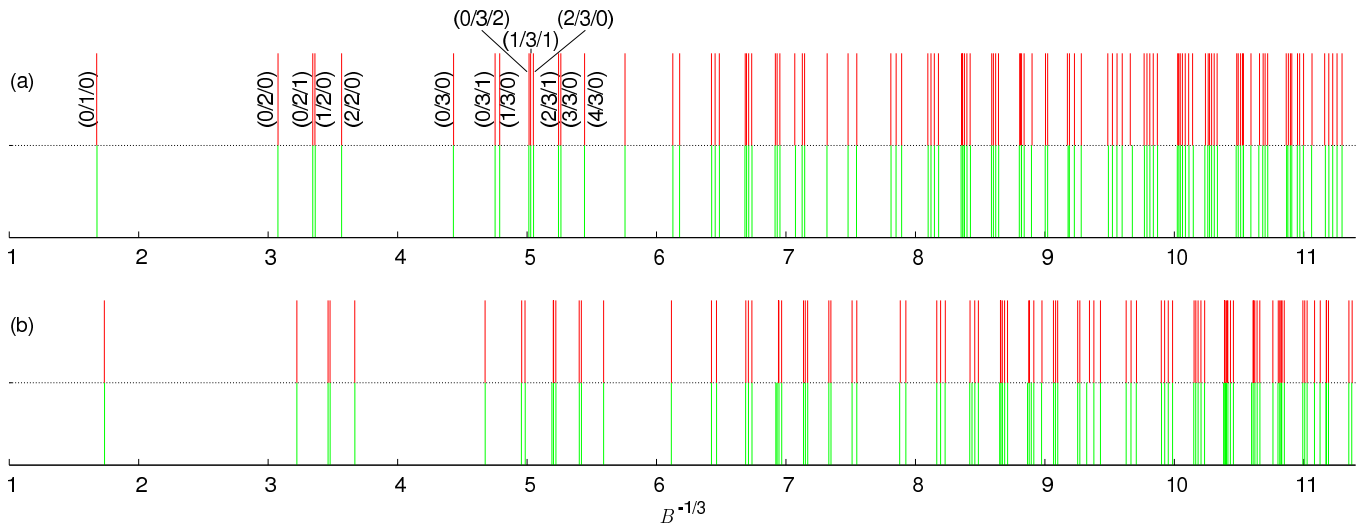}

\caption{(a) The EBK spectrum as obtained from the quantization of the $T_{3}^{\text{p}}$
(upper half, red) at $\tilde{E}=-1.4$ and $\tilde{F}=0.5$ compared to the exact
quantum-mechanical spectrum (lower half, green). The first three $n_{2}$-manifolds
are labeled with ($n_{1}$, $n_{2}$, $n_{3}$). (b) Spectra at $\tilde{E}=-1.5$, $\tilde F=0.5$.}

\label{fig:SCbig}
\end{figure*}

Fig.~\ref{fig:SCbig} shows the EBK spectrum obtained from the quantization
of the $T_{3}^{\text{p}}$. The low-lying states are labeled with
the semiclassical quantum numbers ($n_{1}$, $n_{2}$, $n_{3}$).
The manifolds of constant $n_{2}$, corresponding to the quantization
of $\tilde{I}_{2}$, determine the principal series discernible in
the spectrum. The $n_{2}$-manifolds overlap for $n_{2}>4$. For given
$n_{2}$, $n_{1}$ ranges from 0 to $2n_{2}-2$, giving a total number
of $2n_{2}-1$ subseries. The number of states within one subseries
is $n_{2}-\text{Int}\left[\frac{n_{1}}{2}\right]$. Subseries are
identified by a constant value of $n_{1}+n_{3}$. The states within
one subseries are labeled by $n_{3}$, beginning with 0 for the highest
state. 

Fig.~\ref{fig:SCbig} demonstrates that the spectrum derived from
the quantization of the $T_{3}^{\text{p}}$ is in very good agreement
with the exact quantum spectrum obtained as in, e.g., \cite{ExactQMJoerg,ExactQMTaylor,Ericsonfluctuations}.
The energy levels can be characterized in terms of  three quantum numbers $n$, $q$, $k$, which were first introduced in perturbation theory \cite{Braun}: The principal quantum
number $n$ identifies the principal series. The second quantum number
$q$ runs from $-(n-1),\ldots,+(n-1)$ with increasing energy and
corresponds to the subseries. Finally, $k$ counts the states within
a subseries and runs from $0\:\text{to }n-\left|q\right|-1$. A line-by-line
comparison between the semiclassical and the exact quantum spectrum
as in Table~\ref{tab:SCcorrespondence} yields the following correspondence
between the torus quantum numbers $n_{1},\: n_{2},\: n_{3}$ and the
quantum numbers $n,\: q,\: k$: 
\begin{equation}
\begin{array}{lll}
n & = & n_{2},\\
q & = & (n_{1}+n_{3}+1)-n_{2},\\
k & = & n_{3}.\end{array}
\label{eqn:sccorrespondence}
\end{equation}
 Since $\tilde{I}_{2}$ corresponds to the fastest rotation, it comes
as no surprise that the semiclassical $n_{2}$ represents the principal
quantum number $n$.

\begin{table}
\begin{tabular}{lll|lll|l|l}
 $n$ & $q$ & $k$ & $n_{2}$ & $n_{1}$ & $n_{3}$ &  $B_{\text{qm}}^{-1/3}$ & $B_{\text{EBK}}^{-1/3}$ \\
\hline
1 & 0 & 0 & 1 & 0 & 0 & 1.6773 & 1.6758 \\
 2 & -1 & 0 & 2 & 0 & 0 & 3.0754 & 3.0761 \\
 2 & 0 & 1 & 2 & 0 & 1 & 3.3442 & 3.3448 \\
 2 & 0 & 0 & 2 & 1 & 0 & 3.3632 & 3.3615 \\
 2 & 1 & 0 & 2 & 2 & 0 & 3.5671 & 3.5671 \\
 3 & -2 & 0 & 3 & 0 & 0 & 4.4309 & 4.4320 \\
 3 & -1 & 1 & 3 & 0 & 1 & 4.7528 & 4.7526 \\
 3 & -1 & 0 & 3 & 1 & 0 & 4.7888 & 4.7883 \\
 3 & 0 & 2 & 3 & 0 & 2 & 5.0134 & 5.0150 \\
 3 & 0 & 1 & 3 & 1 & 1 & 5.0279 & 5.0273 \\
 3 & 0 & 0 & 3 & 2 & 0 & 5.0500 & 5.0487 \\
 3 & 1 & 1 & 3 & 2 & 1 & 5.2428 & 5.2437 \\
 3 & 1 & 0 & 3 & 3 & 0 & 5.2619 & 5.2610 \\
 3 & 2 & 0 & 3 & 4 & 0 & 5.4435 & 5.4445 
\end{tabular}

\caption{Semiclassical and exact eigenvalues of the scaling parameter at $\tilde{E}=-1.4$ and $\tilde{F}=0.5$.}

\label{tab:SCcorrespondence}
\end{table}

\begin{figure}
\includegraphics[width=1\columnwidth]{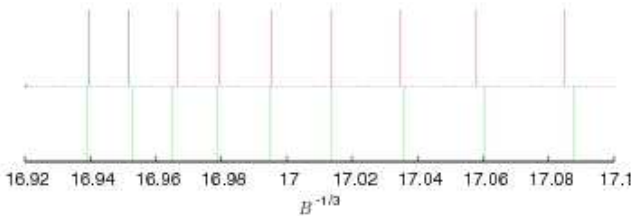}

\caption{The quantization of the $T_{3}^{\text{n}}$ family (blue) gives the
two left-most peaks in the $n_{2}=10$ and $n_{1}+n_{3}=10$ ($n=10$,
$q=1$) subseries showing a good agreement with the exact quantum
spectrum (green). The other peaks result from the quantization of
the $T_{3}^{\text{p}}$ (red). The slight discrepancies of the peaks
on the right are due to numerical inaccuracies in the interpolation
procedure.}

\label{fig:SCneg}
\end{figure}

For large $n_{2}$, the frequency ratios calculated from Eqs.~(\ref{eqn:system1})
and (\ref{eqn:system2}) for the lowest peaks in the central subseries
($n_{1}+n_{3}\approx n_{2}$) lie outside the frequency range covered
by the $T_{3}^{\text{p}}$. These peaks belong to the $T_{3}^{\text{n}}$
family. However, the boundary of frequency ratios, especially in regions
where few POs are available for the calculation of the action variables,
is sometimes not clear-cut, which leads to a fuzzy boundary between
the states of $T_{3}^{\text{p}}$ and those of $T_{3}^{\text{n}}$.
Fig.~\ref{fig:SCneg} shows a spectrum for a single subseries where
the lowest states are taken from the quantization of the $T_{3}^{\text{n}}$.
The quantization of the $T_{3}^{\text{n}}$ requires $\alpha_{1}=0$
and yields a correspondence between torus quantum numbers and traditional
quantum numbers that is different from~(\ref{eqn:sccorrespondence}).
The comparison with the exact quantum spectrum in Fig.~\ref{fig:SCneg}
shows excellent agreement, with the exception of the two right-most
peaks. These states lie in a region of the plane of winding ratios
where not many POs are known and the action variables can be calculated
only to correspondingly lower accuracy.

\section{Conclusion}

\label{sec:conclusion}

The hydrogen atom in crossed electric and magnetic fields presents
long-standing challenges to both the dynamical-systems and the atomic-physics
communities. In this work we have addressed both issues. From the
viewpoint of dynamical systems we have demonstrated the power of periodic
orbits when they are used as a probe to the intricacies of the geometrical
and dynamical structures in phase space. Being the remnants of broken
tori, periodic orbits can be used to establish a complete hierarchy
of these phase space structures. At energies slightly below and slightly
above the ionization saddle point of the crossed-fields hydrogen atom
we established this hierarchy as follows. Three fundamental periodic
orbits \cite{FloethmannKepler} that do not arise from the breakup
of a higher-dimensional torus represent the 1-tori in the hierarchy.
Two of them serve as organizing centers for two families $T_{2}^{\text{p}}$
and $T_{2}^{\text{n}}$ of 2-torus POs. The periodic orbits located
in the $x$-$y$ symmetry plane form the $T_{2}^{\text{p}}$ family. The
2-torus POs themselves were found to be limiting cases of two families
$T_{3}^{\text{p}}$ and $T_{3}^{\text{n}}$ of 3-torus POs. 

Having established the hierarchy of POs, we calculated the individual
action variables for the different families of POs. This knowledge
provides the basis for the semiclassical calculation of the atomic
spectrum using EBK quantization. Our results are in good agreement
with the exact quantum mechanical spectrum. 

Because the classification of POs by winding numbers relies only on the existence of a hierarchy of broken tori, which is a common feature in many non-integrable Hamiltonian systems, it will be applicable to other challenging systems. In particular, previous experience on the hydrogen atom in a magnetic field \cite{Main99b} has shown that a semiclassical approximation that is derived in a near-integrable setting can successfully describe quantum states even deep in the mixed regular-chaotic regime. It is therefore to be expected that our quantization scheme will still be useful at appreciably higher field strengths than were considered here.

\begin{acknowledgments}
We thank C.~Chandre, \`A.~Jorba, J.~D.~Meiss and
G.~Wunner for helpful comments and remarks. This work
was supported by the Deutsche Forschungsgemeinschaft,
Deutscher Akademischer Austauschdienst, National
Science Foundation, and the Alexander von Humboldt-Foundation.
\end{acknowledgments}

\begin{appendix}

\section{The periodic-orbit search}

\label{app:search}

Due to the Coulomb singularity in the Hamiltonian~(\ref{eqn:Hamiltonian}),
a straightforward numerical integration of the trajectories is unfeasible.
To overcome this difficulty, we use the Kustaanheimo-Stiefel regularization
\cite{KS1,KS2} and integrate in \emph{four-dimensional} position and momentum
coordinates $\mathbf{u}$ and $\mathbf{p}$.
The equations of motion in KS coordinates with respect to a pseudotime parameter $\tau$ are
free of singularities (see, e.g., \cite{ExactQMJoerg}).

We describe the search for periodic orbits as a root-finding problem.
A PO is identified by a starting point $P_{i}=\left(\mathbf{u},\mathbf{p}\right)_{i}$
in phase space and by its pseudotime period $\tau$. Given an initial
guess for $P_{i}$ and $\tau$, we calculate the final point $P_{f}$
of the trajectory by integrating the equations of motion for a time $\tau$ and then
use a Newton-like method to modify $P_{i}$ (within the energy shell)
and $\tau$ so that $P_{f}=P_{i}$. Specifically, we use a Powell-hybrid
root-finding method \cite{NAG} for this task. 

The initial guesses required by the root-finder are obtained as follows:
we fix a Poincar\'{e} surface of section in phase space. Initial
guesses for $P_{i}$ are chosen on an equidistant grid on this four-dimensional
surface. Starting from each of these points, we integrate the equations
of motion until the trajectory returns to the neighborhood of its
starting point (or a prescribed maximum period is exceeded). The time
it takes for the trajectory to come back close to its starting point
serves as the initial guess for the period $\tau$. 

Like any other numerical PO search, this algorithm is not guaranteed
to find all POs. It does not require any prior knowledge of the dynamics,
which we extract from the POs \emph{a posteriori}, even if the data
set is incomplete. 

\end{appendix}


\end{document}